\documentstyle[preprint,aps,prd]{revtex}

\input{psfig}
\begin{document}
\draft
\preprint{}
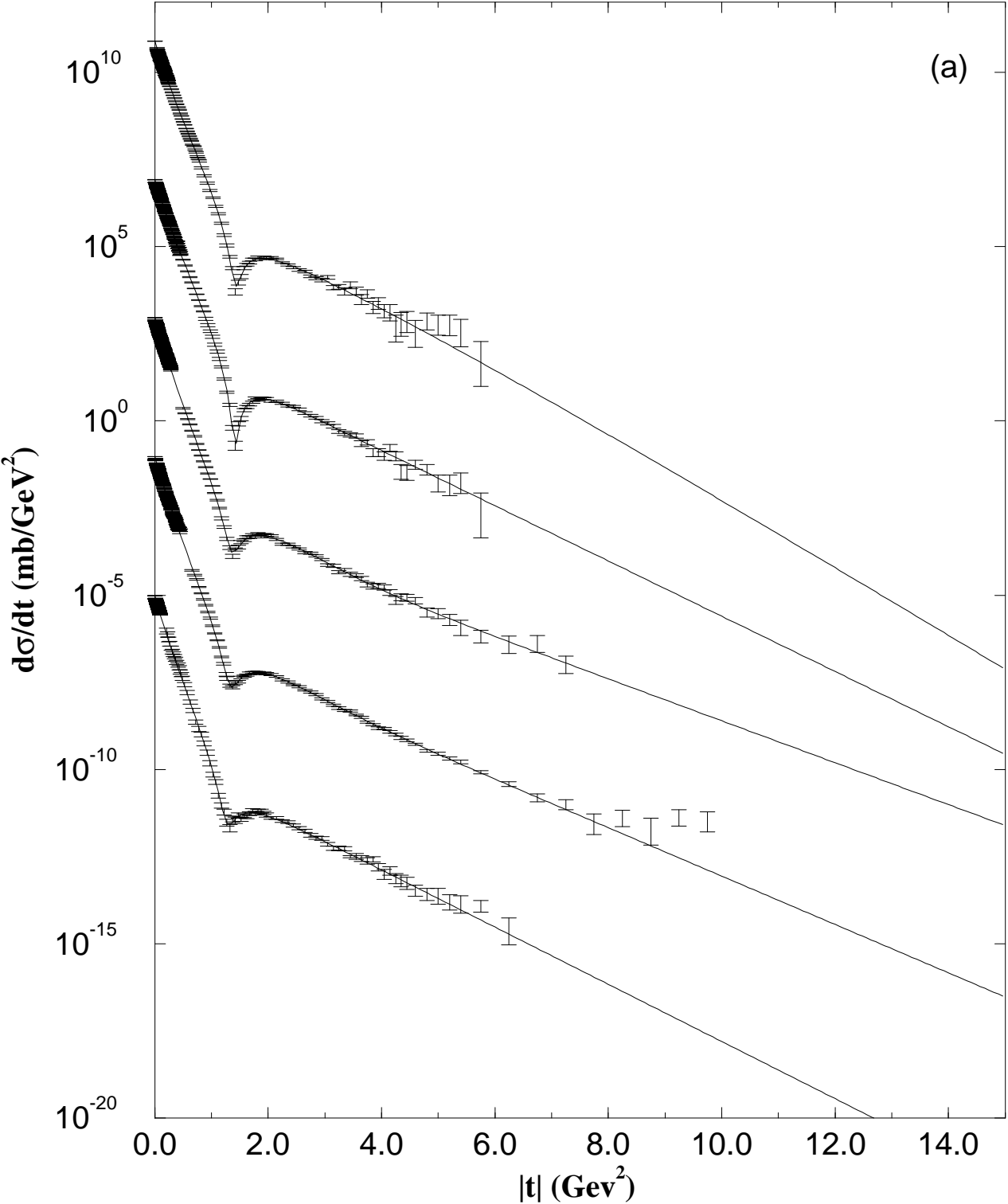
\title{Evidence for eikonal zeros in the momentum transfer space
}
\author{P. A. S. Carvalho and M. J. Menon}
\address{
Instituto de Fisica `Gleb Wataghin'\\
Universidade Estadual de Campinas, Unicamp\\
13083-970 Campinas, SP, Brazil
}
\date{\today}
\maketitle
\begin{abstract}
We present the results of fitting elastic $pp$ differential cross section data at 23.5 $\leq \sqrt{s} \leq$ 62.5 GeV with a novel analytic parametrization for the scattering amplitude. Making use of a fitting method, the errors from the free parameters are propagated to the imaginary part of the eikonal in the momentum transfer space. A novel systematic study of the effects coming from data at large momentum transfer is also performed. We find statistical evidence for the existence of eikonal zeros in the interval of momentum transfer 5-9 $GeV^{2}$.
\end{abstract}

Several authors have investigated elastic $pp$ scattering at high energies in the context of the impact parameter formalism and/or eikonal approximation. Model-independent analyses, performed through fits to differential cross section data, play an important role as a source of empirical information for theoretical developments.

Inspired by some of these model-independent approaches \cite{franco,francahama,sanivalin,lombard,furgetvalin,franca} and based on experimental information presently available \cite{schubert,amaldischubert} we retake two crucial aspects of the empirical analyses: (a) a systematic study of the effect of experimental informations at large momentum transfer; (b) development of a statistical procedure in order to estimate the eikonal uncertainties in the momentum transfer space. In this work we briefly report some results that can contribute with a better understanding of the above aspects, and bring important insights for the development of theoretical approaches.

We shall analyze experimental data on elastic $pp$ differential cross section and the $\rho$  parameter (ratio of the forward real to imaginary part of the scattering amplitude) between $\sqrt{s}=$ 23.5 GeV and $\sqrt{s}=$ 62.5 GeV. Our basic set corresponds to the data compiled and normalized by Amaldi and Schubert at the energies $\sqrt{s}=$ 23.5, 30.7, 44.7, 52.8, and 62.5 GeV and momentum transfer in the interval 0.01 GeV$^{2} \leq q^{2} \leq$ 9.8 GeV$^{2}$ \cite{schubert,amaldischubert}. However, based on the evidence that these differential cross section data do not depend on the energy for momentum transfer above $q^{2} \sim$ 3.0 GeV$^{2}$ \cite{donnachielandshoff,faissler} we shall also take account of the data available at $\sqrt{s}=$ 27.5 GeV, in the interval 5.5 $\leq q^{2} \leq$ 14.2 GeV$^{2}$ \cite{schubert}.

Our strategy in performing a systematic study of this experimental information is to consider two different ensembles of data, characterized and denoted by the following.
\begin{itemize}
\item Ensemble $A$

Five original sets of experimental data as compiled by Amaldi and Schubert at $\sqrt{s}=$ 23.5, 30.7, 44.7, 52.8, and 62.5 GeV \cite{schubert,amaldischubert}.
\item Ensemble $B$

Five sets of ensemble $A$ including at each energy the data at $\sqrt{s}=$ 27.5 GeV \cite{schubert}.
\end{itemize}

We want to investigate the effects in the eikonal (transfer momentum space) coming from the fits to both ensembles.

We make use of the following analytical parametrization for the scattering amplitude \cite{menoneeu}:

\begin{eqnarray}
F(q,s)&=&i \sum_{j=1}^{n} \alpha_{j} e^{-\beta_{j} q^{2}} - \mu \sum_{j=1}^{2} \alpha_{j} e^{-\beta_{j} q^{2}},\nonumber\\
\\
\label{param}
\mu&=&{-\rho(s) \over \alpha_{1} + \alpha_{2}} \sum_{j=1}^{n} \alpha_{j},\nonumber
\end{eqnarray}
where $\alpha_{j}, \beta_{j}$, $j=1,2,...,n$ are real free parameters and $\rho(s)$ is the experimental $\rho$  value at each energy \cite{amaldischubert}. With this amplitude, we analyze the differential cross section

\begin{equation}
{d\sigma \over dt}= \pi |F(q,s)|^{2}, \qquad -t=q^{2}.
\label{scdif}
\end{equation}

Fits to experimental data from ensembles $A$ and $B$ were performed through the CERN-MINUIT routine, which gives also the errors $\Delta \alpha_{j}$, $\Delta \beta_{j}$ in the free parameters \cite{james}. In this procedure all the parameters were completely free, taking different values at each energy. With this, no assumption at all was made on the dependences of the free parameters with energy or momentum transfer, allowing a good statistical result: $1.0\leq \chi^{2}/N_{DF}\leq2.1$ for a number of $N_{DF}$ between 116 and 226, in all  cases \cite{menoneeu2}. Propagation of the errors to the differential cross section \cite{bevington} takes account of extrapolated curves, which, as stressed by Lombard \cite{lombard}, cannot be excluded on statistical grounds. The complet fit results for ensembles $A$ and $B$ are displayed in Fig.~\ref{fig2} \cite{menoneeu2} (numerical values of the free parameters are available from the authors).

The complex eikonal in the momentum transfer space, $\chi(q,s)$, is connected with the scattering amplitude, Eq. (1), through the well-known formulas

\begin{equation}
F(q,s)=i \int b db J_{0}(q,b) \Gamma(b,s) \equiv i \langle \Gamma(b,s) \rangle,
\label{amp}
\end{equation}

\begin{equation}
\Gamma(b,s)=1-e^{i\chi(b,s)},
\label{perfil}
\end{equation}

\begin{equation}
\chi(q,s)=\langle\chi(b,s)\rangle,
\label{eic}
\end{equation}
where $J_{0}$ is the Bessel function (azimuthal symmetry assumed), $\Gamma(b,s)$ is the profile function and the angular brackets denote a symmetrical two-dimensional Fourier transform.

Here we are only interested in the imaginary part of the eikonal, $\chi_{I}$, which from Eq. (\ref{perfil}) (impact parameter space) reads

\begin{equation}
\chi_{I}(b,s)=\ln{1 \over \sqrt{\Gamma^{2}_{I}(b,s)+[1-\Gamma_{R}(b,s)]^{2}}}.
\label{eicimag}
\end{equation}
The complex profile $\Gamma_{R}(b,s)+i\Gamma_{I}(b,s)$ is analytically determined through Eqs. (1) and (\ref{amp}), together with the corresponding errors by propagation. Since we found that, even taking account of the errors,
\begin{equation}
{\Gamma_{I}^{2}(b,s)\over[1-\Gamma_{R}(b,s)]^{2}}\ll1,
\label{assint}
\end{equation}
the eikonal may be expressed by

\begin{equation}
\chi_{I}(b,s)\approx\ln{1\over1-\Gamma_{R}(b,s)}.
\label{eicaprox}
\end{equation}
From Eqs. (1) and (\ref{amp}), this quantity may be evaluated in terms of the fit parameters and also the corresponding errors, $\Delta\chi_{I}$, by propagation.

At last we should evaluate the eikonal in the momentum transfer space, Eq. (\ref{eic}), and the errors $\Delta\chi_{I}(q,s)$. However, due to the structure of our parametrization (1), the transform (\ref{eic}) cannot be analytically performed and so the errors cannot be estimated as done in the previous steps. We recall that this is a typical problem in current empirical analysis.

In order to evaluate the erros $\Delta\chi_{I}(q,s)$ we introduce the following method. Generically we can expand Eq. (\ref{eicaprox}) in the form

\begin{equation}
\chi_{I}(b,s)=\Gamma_{R}(b,s)+D(b,s),
\label{expansao}
\end{equation}
where $D(b,s)$ corresponds to the remainder of the series. Performing the Fourier transform we obtain

\begin{equation}
\chi_{I}(q,s)=F_{I}(q,s)+D(q,s).
\label{expansaoq}
\end{equation}
Since the amplitude $F_{I}$(q,s) and errors $\Delta F_{I}(q,s)$ are directly given by the fits, our task concerns the evaluation of 

\begin{equation}
D(q,s)=\langle D(b,s) \rangle,
\label{Dtransf}
\end{equation}
with the corresponding errors, $\Delta D(q,s)$, and this is the central point of the method.

First, from Eqs. (\ref{eicaprox}) and (\ref{expansao}), the quantity $D(b,s)$ may be evaluated

\begin{equation}
D(b,s)=\ln[{1\over1-\Gamma_{R}(b,s)}]-\Gamma_{R}(b,s),
\label{D}
\end{equation}
and also the errors, $\Delta D(b,s)$, through error propagation from $\Gamma_{R}(b,s)$. Next, making use of the CERN-MINUIT routine this set of points with errors, $D(b,s)$ $\pm$ $\Delta D(b,s)$, was fitted by a sum of Gaussians

\begin{equation}
D_{f}(b,s)=\sum_{j=1}^{6} A_{j}e^{-B_{j}b^{2}}.
\label{dparamet}
\end{equation}
With this parametrization, $D(q,s)$ in Eq. (\ref{Dtransf}) may be analytically evaluated and also the errors, $\Delta D(q,s)$, may be estimated through the propagation of the errors in $A_{j}$, $B_{j}$, as given by the routine. At last, Eq. (\ref{expansaoq}) leads to $\chi_{I}(q,s)$ and the error propagation furnishes $\Delta\chi_{I}(q,s)$.

This method was used by Furget in order to determine the eikonal $\chi_{I}(q,s)$ \cite{furget}. A novel aspect of our approach is its use in the estimation of errors.

Two typical results, at $\sqrt{s}= 30.7$ and $\sqrt{s}= 52.8\ GeV$, are displayed in Figs.~\ref{fig3} and \ref{fig4} for ensembles $A$ and $B$, respectively. We plotted the quantity $\chi_{I}(q,s)$ multiplied by $q^8$, $q^2=-t$ \cite{furgetvalin}. In the case of ensemble $A$, experimental data at $\sqrt{s}= 30.7\ GeV$ are available up to $t_{max}= 5.75\ GeV^2$ and from Fig.~\ref{fig3} we see that no zero can be inferred. However, from Fig.~\ref{fig4}, the data at large $q^2$ (ensemble $B$, with $t_{max}=14.2\ GeV$), lead to statistical evidence (with the uncertainties) for the existence of zero. This effect of the data at large momentum transfer is also evident at $\sqrt{s}= 52.8\ GeV$ (Figs.~\ref{fig3} and \ref{fig4}) since in ensemble $A$ experimental data at this energy are availiable up to $t_{max}= 9.75\ GeV^2$.

We conclude that data at large momentum transfer allow the identification of eikonal zeros (change of sign). Quantitatively the position of the zero and its uncertainties may be extracted from the plots as shown in Figs.~\ref{fig3} and \ref{fig4}. The results from ensembles $A$ and $B$ are displayed in Fig.~\ref{fig7}. Due to the uncertainties and, meanly, its asymmetry around some points, the dependence of the position of the zero with the energy is difficult to be inferred on statistical grounds.

However, if we disregard the asymmetries and take the uncertainty at larger momentum transfer symmetrically around each point, we are lead to the following quantitative estimates. Parametrizing the results from ensemble $B$ (Fig.~\ref{fig7}(b)) by
\begin{equation}
q_{0}^{2}=A+B\ln{s}
\label{linear}
\end{equation}
gives $A=10.66\pm3.75\ GeV^2$, $B=-0.9917\pm0.993\ GeV^2$ with $\chi^{2}/N_{DF}=1.27$. Although this indicates some decreasing with the energy, the slope is nearly compatible with zero. Assuming zero slope we obtain $q_{0}^{2}=6.9\pm0.7$ GeV$^2$ with $\chi^{2}/N_{DF}=2.07$.

By the other hand, taking account of all the uncertainties displayed in Fig.~\ref{fig7}(b), we could roughly estimate the position of the zero at $7.0\pm2.0$ GeV$^2$ in the interval 23.5 GeV $\leq \sqrt{s} \leq$ 62.5 GeV. 

Extensions of our analysis at $\sqrt{s}=$ 13.8 and 19.4 GeV  are in course and this will be crucial for extracting the dependence of the zero with the energy. Also, detailed discussions on physical interpretations, connections with nonperturbative QCD, phenomenological approaches and comparisons with other analyses are being concluded.

\acknowledgments

We thank the Funda\c c\~ao de Amparo \`a Pesquisa do Estado de S\~ao Paulo - FAPESP for financial support.

\begin{figure}
\caption{Fits to pp differential cross section data through Eqs. (1) and (\protect\ref{scdif}) between 23.5 GeV (upper) and 62.5 GeV (down): (a) ensemble A; (b) ensemble B. Curves and data were multiplied by factors of \protect10$^{\protect\pm 4}$ \protect\cite{menoneeu2}.}
\label{fig2}
\end{figure}

\begin{figure}
\caption{Extracted imaginary part of the eikonal multiplied by t$^4$ from ensemble A: (a) $\protect\sqrt{s}=$30.7 GeV; (b) $\protect\sqrt{s}=$ 52.8 GeV.}
\label{fig3}
\end{figure}

\begin{figure}
\caption{Same as Fig.~\protect\ref{fig3} from ensemble B.}
\label{fig4}
\end{figure}

\begin{figure}
\caption{Position of the zeros in the eikonal as function of the energy from ensemble A (a) and ensemble B (b).}
\label{fig7}
\end{figure}

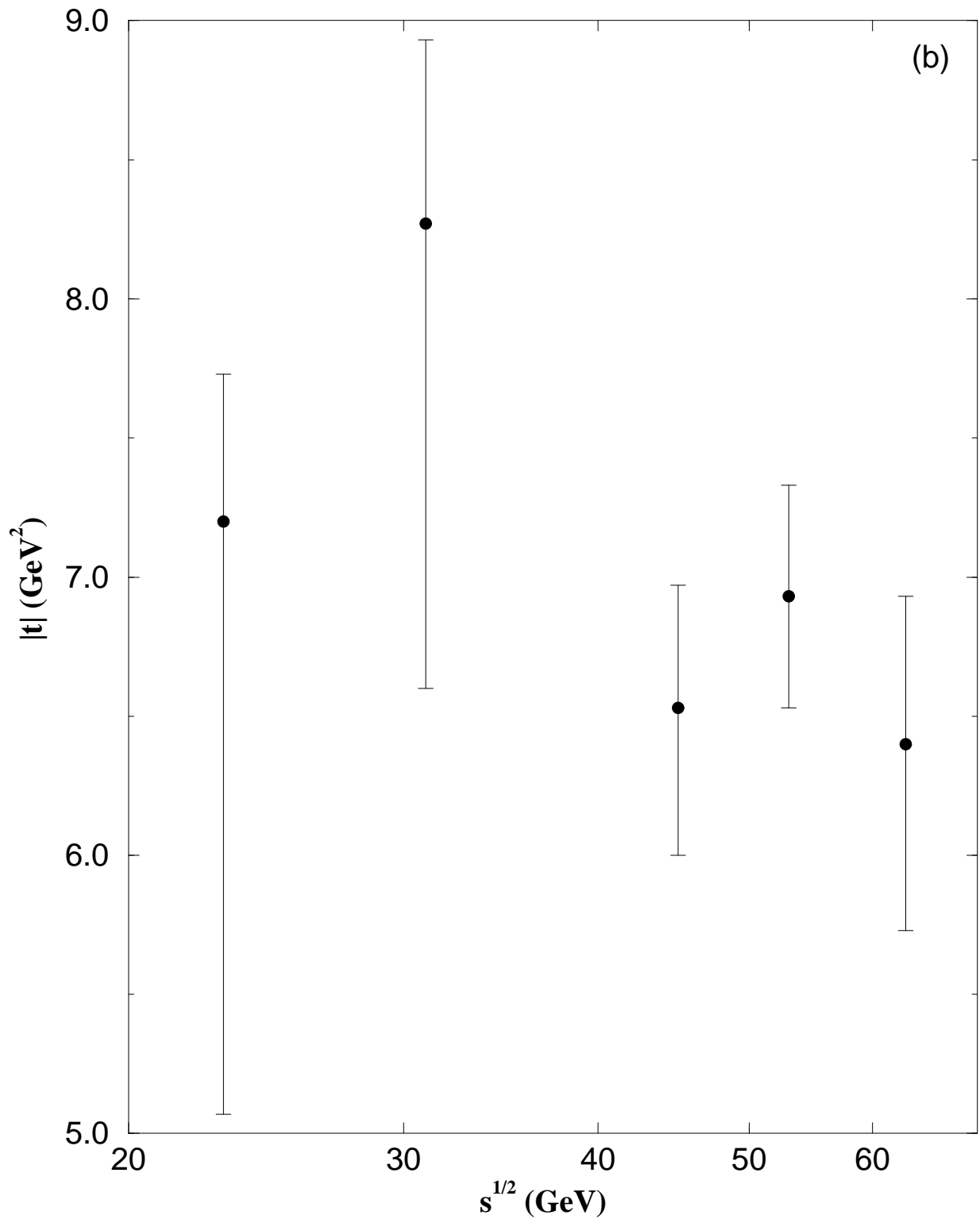
\end{document}